# Elastic Nanocomputation in an Ideal Brain <span style="font-size:smaller">(1p abstract + 36 pages + 49 endnotes)</span>

William Softky, Ph.D, Physics, California Institute of Technology

(not currently affiliated)

## Abstract

This explanation of what a brain is and does rests on informational first principles, because information theory, like its parent theory thermodynamics, is mathematically sacrosanct, itself resting on real-valued probability. Just as thermodynamics has enabled hyper-potent physical technologies from the internal combustion engine to the hydrogen bomb, so information theory has enabled hyper-persuasive technologies, from color television to addictive video games. Only a theory of what a brain is and does based on those same principles makes legible and transparent the mechanisms by which such hyper-persuasion works. In information-theoretic terms, a brain is a specialized real-valued real-time 3-D processor detecting discontinuities in spacetime outside itself and reconstituting in itself a continuous reality based on them. This continuous approach is difficult to reconcile with any computational architecture based on separate neurons, and in fact the vast discrepancy in efficiency (of order at least $10^8$) between those efficiencies constitutes this paper's calculations. This remarkable signal-processing requires strong prior hypotheses embedded in 3-D edge-detecting algorithms, priors which unfortunately also open an unpatchable security hole to automated persuasion. So a 3-D model of the brain is essential for understanding how and why persuasive technologies alter our perception of reality, and for protecting us against systemic, systematic cognitive manipulation.

This paper proposes that the laws governing brains and bodies are at root as simple as Newton's laws of motion for hard-edged objects, because the brain must represent the laws. Specifically, this paper infers and quantifies the elegant physical and computational substrate of an ideal brain, a "frictionless" computer specialized for body control. Physically, an ideal brain would look like a real one—requiring a skull, constant body temperature, neurons, and action potentials—while additionally interwoven with a specialized but nearly undetectable nanoscopic mesh. Computationally, an ideal brain is a multiscale Newtonian simulator and controller, representing continuous space and time. Here we show that an ideal brain performs orders of magnitude better on standard computational measures such as representational capacity, energy efficiency, and speed than circuits composed only of neurons. An ideal brain matches human performance more closely, but only in some ways, suggesting a latent reservoir of "dark computation" and apparently super-natural (but necessarily sub-conscious) abilities. Reviving this circuitry requires treating brains as specialized real-time signal-processors rather than as general-purpose thinking machines.[1]

## The Ideal Brain

Most mammalian brains move tree-structured quadruped bodies, which although mechanically stable still require precise, continuous control in space and time, along with ongoing self-calibration. Quadruped locomotion already challenges current technological limits[2], but homonid brains faced an even harder task demanding yet more precision: balancing an unstable upright structure (physically, a seven-layer inverted compound pendulum) well enough to stand, walk, and run without wasting metabolic energy. A later section of this paper, *Continuous Control of Millions of Muscles*, outlines the informational principles by which even a toddler can learn to walk.

Walking does not require the usual "digital" features of quantized, persistent storage or categorical decisions[3]. But nonetheless quantized states did evolve in brains, as evidenced by episodic memory, language and storytelling. It seems likely the multiscale quickness required for bipedal locomotion eventually also contributed to *homo sapiens'* categorical thinking prowess[4].

Dynamic balance of any structure, much less an unstable one, requires simulating a set of continuously-evolving differential equations, a task whose tight constraints and specificity cry out for specialized analog hardware. Additionally, circuit stability must depend on copious high-resolution data about the current physical world at widely varying timescales, and the equivalent of continuously self-correcting closed-loop control equations valid at arbitrary levels of resolution. While such ever-evolving, ever-decaying analog computation is ill-suited to execution on digital hardware, the reverse is not true: analog substrates can and do support quantized computation[5].

So it is possible that a biped's demand for tighter spatiotemporal precision in the continuous domain enabled *homo sapiens* to grow a novel, *quantized* form of storage atop the continuous substrate: persistent, self-reinforcing states and narrative structures, such as oscillating loops, which could underlie the unique mental abilities humans seem most proud of. This would give us a hybrid system made of two interleaved operating systems, continuous real-time and quantized, each running at many timescales simultaneously. The final section of this paper, *Instabilities of a Hybrid Real-time/Episodic System*, describes the dangers of too powerful a mind.

Speculating on such topics does not fit within the body of this paper. This work instead approaches a far simpler problem at the opposite end of the representational spectrum, an idealization of the primary computational task facing every mammalian brain: how to represent the three-dimensional world as accurately and efficiently as possible at the local scale[6]. The guiding principle for such simulation (best articulated by Carver Mead[7]) might best be called "resonant computing": that the most efficient simulation of a physical system should match the target system at the finest level possible.

Unfortunately, any network composed of discrete neurons has a very different topology from the truly continuous body-control problem it must solve. I do not claim neural networks *cannot* solve that problem, but that they do so millions-fold less efficiently than is physically possible. That claim implies comparison. So as a "reference idealization" for the entire body-control problem (which includes sensorimotor feedback) I choose only its most resource-hungry subcomponent, the part representing 3-D Newtonian space, and within that subcomponent I discuss only its use of physical resources like energy, space, and time. Furthermore, as a reference idealization for the optimum, "frictionless" computational medium against which a neuron-only model might compete, I hypothesize *simulatrix* (outlined in the first section below), a physically plausible, nanoscopic matrix[8] interpenetrating and interacting with neocortical neurons to help them simulate the outside world.

The following sections provide order-of-magnitude advantage ratios by which the performance of an ideal neurons-plus -simulatrix brain surpasses a brain using neurons alone. The ratios are as follows:

| Learning-speed advantage for 3-D priors | Eqn. 3 | $10^3 - 10^9$ |
|---|---|---|
| Settling-speed advantage for synchronized spikes | Eqn. 8 | $10^4 - 10^{11}$ |
| Power advantage for non-volatile static memory | Eqn. 11 | $10^5$ |
| Power advantage for non-volatile momentum memory | Eqn. 14 | $10^3 - 10^9$ |
| Energy advantage for diffusive smoothing | Eqn. 18 | $10^5 - 10^{15}$ |
| Memory-capacity advantage for tininess | Eqn. 19 | $10^9$ |
| Power advantage for native floating computation | Eqn. 22 | $10^{12} - 10^{20}$ |
| Bandwidth advantage for vibratory communication | Eqn. 25 | $10^4$ |

These advantages are estimated separately below. To the degree that they represent redundant fitness factors, their geometric mean (about $10^8$) could approximate the net advantage of simulatrix. If on the other hand the advantages represent independent fitness factors, then the advantage of simulatrix might be yet larger, up to $10^{64}$. Both discrepancies are enormous, but there is in fact a surprisingly simple reason they might have been missed.

## Simulatrix

Computers must benefit from nanotechnology[9], because smaller circuit elements process more information, and even faster. The same reasoning applies to brains. So an ideal brain must compute at the most granular possible resolution with the least possible energy, and its elemental data format—its continuous computational substrate, simulatrix—must be locally isomorphic to the outside world. In fact, simulatrix represents outside 3-D space by itself *being* a 3-D medium. Furthermore, as an elastic 3-D mesh akin to an aerogel, it supports longitudinal elastic waves (P-waves of variable velocity) to represent motion [Figure 1].

Non-uniformities in simulatrix, whether static kinks or moving waves, store and represent scalar 3-D probability information: the positions of object contours, the current orientation of a skeletal joint (elevation, azimuth, twist), or 3-D parameterized abstractions of arbitrary sensory input. As physical surfaces in a 3-D gel, those probability contours may propagate with little dispersion or damping, so their own momentum could represent outside momentum. In simulatrix, moving P-waves are the native representation of linear change across space and time.

The "input" to simulatrix is tiny, point-like synaptic events, at the sub-millisecond and sub-micron scale of synaptic vesicle release. Those impulses initiate small spherical P-waves whose collective correlations and coherence learn, represent, and ultimately reconstitute the structure of the world outside by means of overlapping spherical ripples (as from raindrops in a 3-D pond) and occasional large waves. The physical residue of interference fringes can accumulate as subsequent 3-D priors, like in holograms.

Simulatrix must be a passive medium, making no choices or energetic investments beyond moving P-waves triggered by synaptic events and accumulating probabilities. But even an ideal brain must choose how to spend its limited energy budget, and this complementary task is performed by neurons. When a sufficiently large P-wave passes through a neuron, the transient non-uniformity in simulatrix boosts dendritic depolarization enough to detect and relay that wavefront's passing by means of an action potential. Several nearby neurons in parallel, all triggered by the same P-wave, could reconstitute or "teleport" the wavefront elsewhere via a coherent volley of spikes and subsequent synaptic events[10].

While clearly speculative, this scheme accounts for certain features of real brains. For example, an elastic computer would need vibrational protection, which a skull could provide. A single neuron triggered by uncorrelated passing waves would fire irregularly, as real neurons do[11]. And because wavefront teleportation requires consistent phase, axonal conduction velocity and thus brain temperature must be tightly regulated.

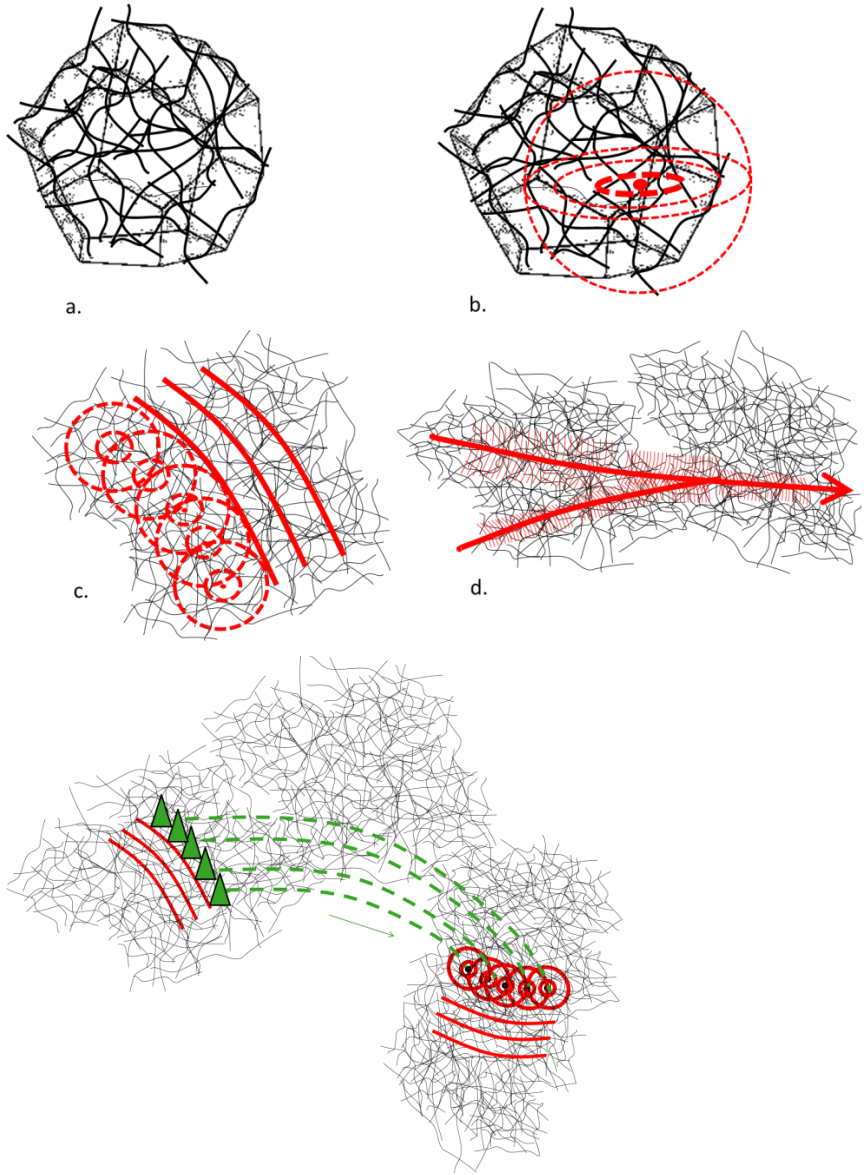

Figure 1: a) Simulatrix is a molecular-scale mesh filling neocortical space which computes continuously in space and time. b) A synaptic event triggers a single elastic pulse like a spherical blast-wave, which coasts outward using no additional power. c,d ) Several synaptic events in synchrony create traveling compression waves whose motion represents probability contours ("P-waves"). e) P-waves can be teleported elsewhere by neurons firing in synchrony, using Huygens' superposition principle.

## An ideal brain must be specialized for 3-D computation

Is a human brain a generic, general-purpose computational system, or a special-purpose one? Certainly brains have generic properties. They fuse spiking input across sensory modalities, and they can learn to perceive or control almost anything. Because neocortex has the same basic laminar structure and connectivity profile in both sensory and motor areas, it is tempting to assign it a general algorithmic function, such as a "canonical micro-circuit" or a Hopfield net which relaxes over time into a solution state[12].

A brain's initial and most crucial task is quite specific: to model the body's shape, that is, to associate specific motor and sensory responses (say, spikes from muscle-strain receptors) with specific joint positions, and to generalize them into a smooth and self-consistent motor map. That takes time, even in the ideal case that the essential priors of Newtonian dynamics are pre-wired into the circuit rather than learned[13]. [Figures 2, 3]

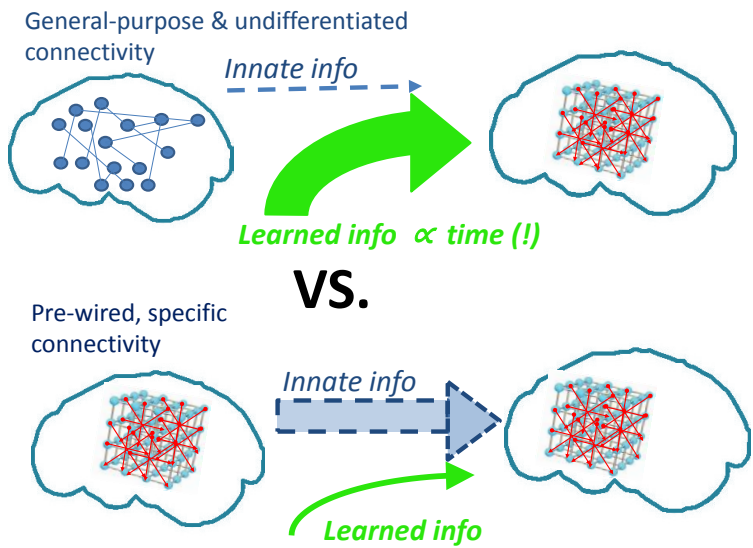

Figure 2: The information necessary to select a 3-D spatial arrangement of voxels must be learned if it is not innate.

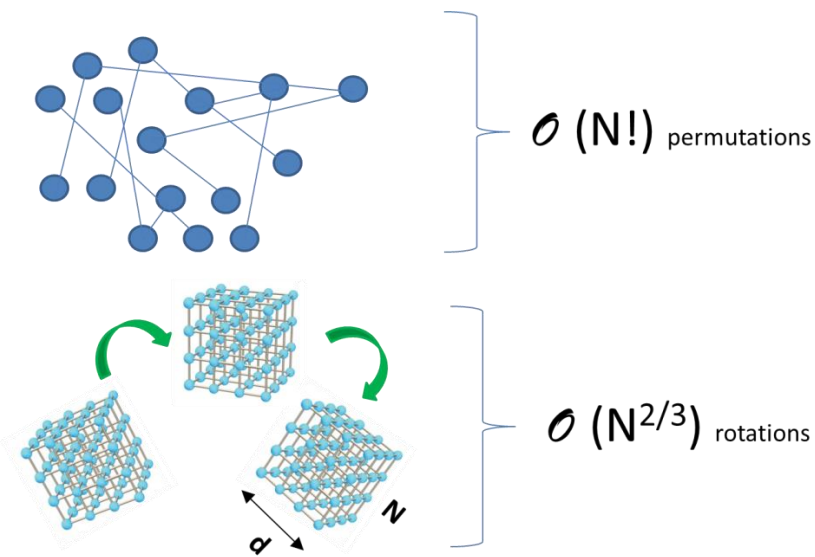

Learning-speed advantage for 3-D priors
= $\log_2(N!) / \log_2(N^{2/3})$ ~ $N$ ~ $10^3 - 10^9$

Figure 3: An already-ordered cube has $N^{2/3}$ orientations from which to select, while there are far more permutations (N!) of the same voxels. The information ratio of the two cases is approximately N.

How much time would pre-wiring save? To calculate comparison here, as subsequently, we approximate a block of (continuous) simulatrix as a block of **N** distinct voxels (volume units). An example of pre-wired Newtonian structure would be a 3-space cube containing **N** voxels stacked like oranges in a box (coarse=10x10x10=$10^3$, fine=1000x1000x1000=$10^9$), with near-neighbor connections already established. Since only the cube's overall orientation is left to be initialized, sensory input merely has to disambiguiate among those orientations. That number of orientations scales not with the number of voxels in the cube (the volume) but with its surface area, I.e. **N** $^{2/3}$ :

$$\textbf{info}_{\textbf{prewired}} \quad \alpha \ \log_2(\textbf{N}^{\ 2/3}) \tag{1}$$
$$\cong \ 2/3 \log_2 (\textbf{N})$$

That's the pre-wired case. On the other hand, a general-purpose connectionist arrangement of those same voxels would be unconstrained *ab initio*. That is, the unordered voxels are not pre-wired into a 3-D lattice, so their permutations are now **N!**, and the minimum information one must accumulate to register them in an oriented cube (even in the noise-free case!) is now

$$\textbf{info}_{\textbf{unwired}} \quad \alpha \ \log_2(\textbf{N!}) \tag{2}$$
$$\cong \ \textbf{N} \log_2(\textbf{N}) - \textbf{N}$$
$$\cong \ \textbf{N} \log_2(\textbf{N})$$
$$\cong \textbf{10}^3 \log_2 (\textbf{N}) \text{ (coarse)}$$
$$\cong \textbf{10}^9 \log_2 (\textbf{N}) \text{ (fine)}$$

*advantage for 3-D priors* (3)
$$= \textbf{info}_{\textbf{unwired}} / \textbf{info}_{\textbf{prewired}}$$
$$= \textbf{10}^3 \text{ (coarse)}$$
$$= \textbf{10}^9 \text{ (fine)}$$

So learning three-dimensionality from experience takes from a thousand to a billion times more data (and hence more time) than having it already built-in. This result makes sense because optimum signal-processing by definition exploits all the prior information that it can, and three-dimensionality is the only certain fact a brain could ever know[14].

Besides learning speed, the other tight constraint is execution speed: upright balance demands that signals must be acted on very, very quickly with precise phase. In simulatrix, P-waves are driven by and read out as synchronized volleys of precisely-timed spikes[15]. If each spike has an inherent phase precision of about a millisecond, and the average phase-jitter of the volley improves as $N^{-1/2}$, then:

$$T_{synch} \text{ (coarse)} = \frac{1 \text{ msec}}{\sqrt{10^3}} = 30 \text{ usec} \qquad (4)$$

$$T_{synch} \text{ (fine)} = \frac{1 \text{ msec}}{\sqrt{10^9}} = 10^{-7.5} \text{ sec} = 0.03 \text{ usec}$$

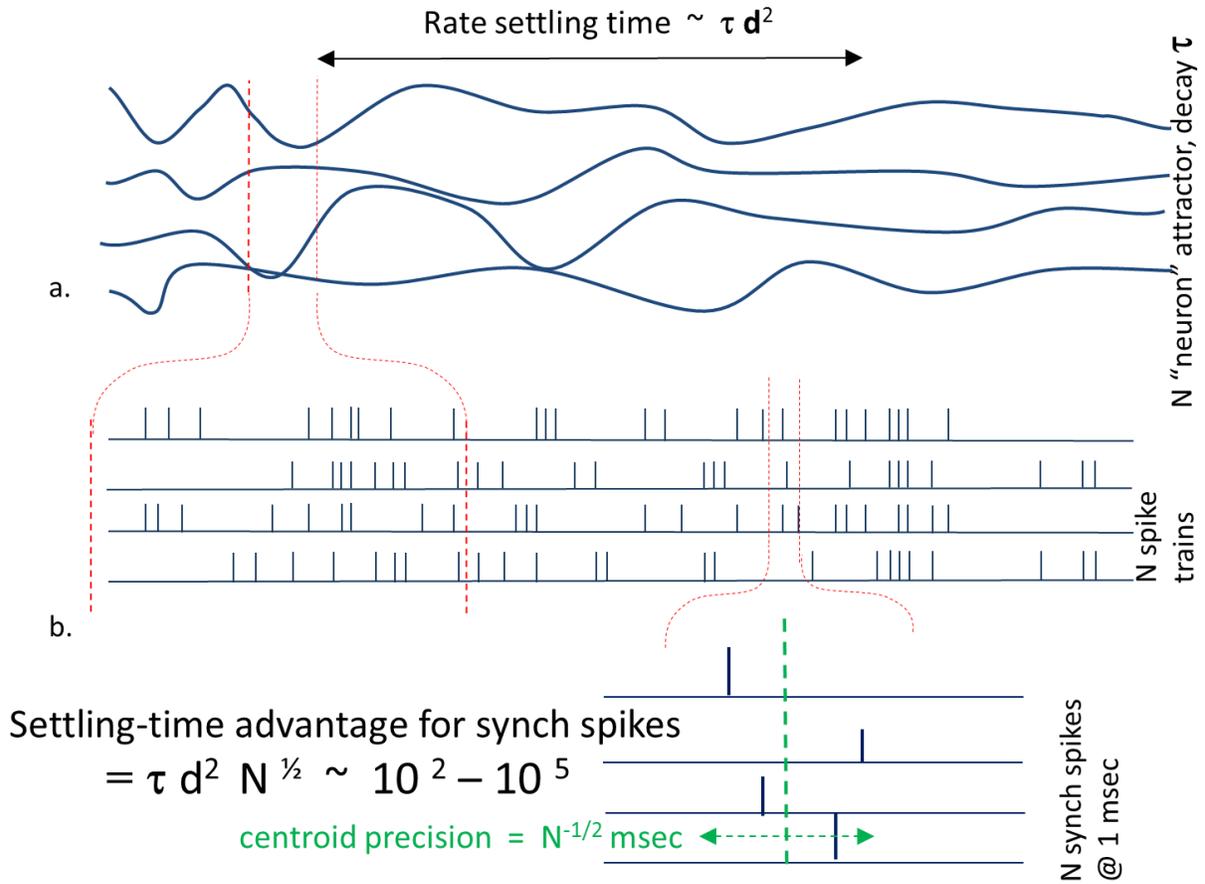

Figure 4: Spike trains can be interpreted as either rate-codes or synchrony codes. (a) The temporal precision of rate-coded attractor networks is limited by the single-unit decay and collective diffusion (which worsens with $d^2$). (b) Synchrony codes are limited instead by the faster single-spike resolution and volley precision, which improves as $N^{-1/2}$.

Connectionist networks, on the other hand, are structurally slower because they relax over time as information diffuses through them. At minimum a network's collective settling-time (and hence effective "phase precision") will be limited by the time constant $\tau$ of a single unit:

$$\mathbf{T_{diff}} \geq \tau \tag{5}$$
$$\geq \mathbf{10\ msec\ (typical)}$$

That's just the minimum, without diffusion. But if units are connected only locally, the settling-time increases with the square of the minimum number of hops connecting typical units, because in diffusion distance scales with time as

$$\mathbf{T} \propto \mathbf{d}^2 \tag{6}$$

For example, a modest $\mathbf{N}=10^3$ cube has $\mathbf{d}=10$ hops between opposite corners, and a $\mathbf{N}=10^9$ cube $\mathbf{d}=1000$, so the diffusion time is:

$$\mathbf{T_{diff}}\text{ (coarse)} \propto \mathbf{10\ msec * 10^2} = \mathbf{1\ sec} \tag{7}$$
$$\mathbf{T_{diff}}\text{ (fine)} \propto \mathbf{10\ msec * 1000^2} = \mathbf{10^4\ sec}$$

*advantage for synchronized spikes* (8)
$$= \mathbf{T_{diff}} / \mathbf{T_{synch}}$$
$$= \mathbf{1\ sec\ /\ 30\ usec} \cong \mathbf{10^4}\ \mathbf{(fine)}$$
$$= \mathbf{10^4\ sec\ /\ 10^{-7.5}\ sec} \cong \mathbf{10^{11}}\ \mathbf{(coarse)}$$

In summary, a representation made of simulatrix can learn a million-fold faster and operate at least ten-thousand-fold more precisely than one made of neurons.

## An ideal brain does not waste energy on anything predictable

In a brain, as in portable electronics, expensive energy should not be wasted preserving information which doesn't change. Electronics solve this problem with "non-volatile memory," such as magnetic domains or "flash" silicon memory (like in thumb-drives), which once set can hold state for years without additional power.

Simulatrix operates on the same principle, because each synaptic event creates a P-wave which carries at least one bit of information and survives indefinitely (seconds, hours, maybe years) without refreshing. So simulatrix' power consumption is strictly capped:

$$P_{simulatrix} \ll 1 \text{ spike/(bit-sec)} \qquad (9)$$

Connectionist models, on the other hand, have two forms of memory. The most durable and fine-grained synaptic memory is non-volatile, like simulatrix, since synapses presumably need no power to preserve their strengths. But in connectionist attractor models, active "state" is maintained by active neural firing, costing energy [Figure 5]. One simulated network of $2*10^6$ neurons firing (presumably at more than 1 Hz) can hold a few digits' worth of information[16], ultimately using 2 million spikes in order to hold a few dozen bits, I.e. $10^5$ spikes/(bit-second). A more expensive simulation yielded similar values[17] by activating an entire cortical column to store a single bit, meaning that for both simulations

$$Power_{attractor} \sim= 10^5 \text{ spikes/(bit-sec)} \qquad (10)$$

*advantage for non-volatile static memory* $\qquad (11)$
$= Power_{attractor} / Power_{simulatrix} \sim 10^5$

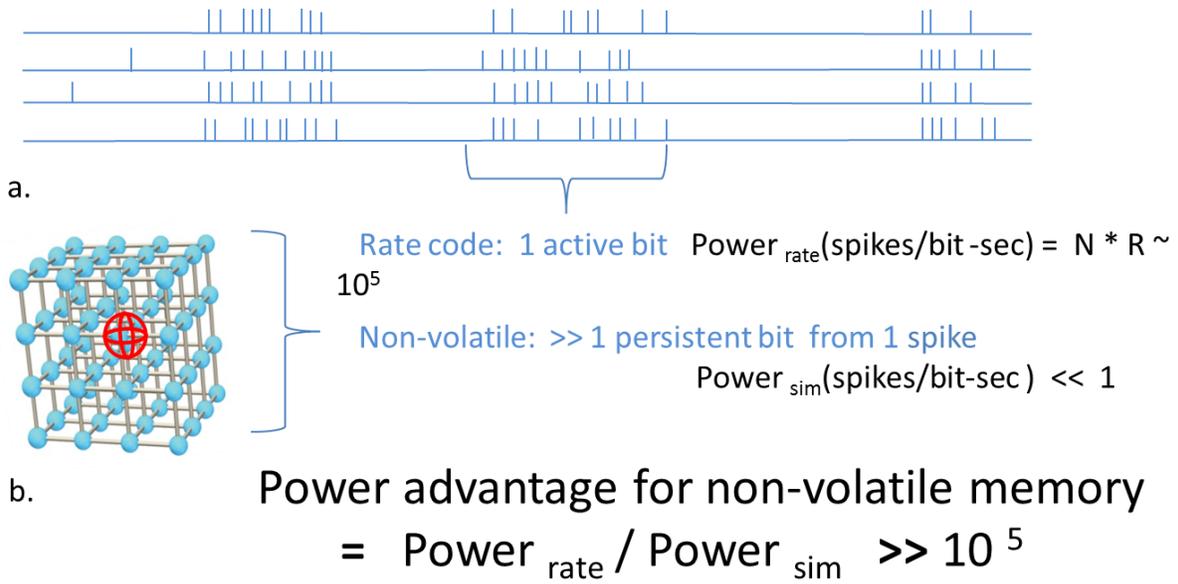

Figure 5: (a) An attractor state using a rate code consumes power proportional to both rate and the neural population. (b) In simulatrix, many bits can be set and maintained quasi-permanently by a single synaptic event, so the ratio of the two is of order NR.

Simulatrix has a further benefit not seen in dissipative circuits: the ability to hold a *changing* state without dissipating power, via its use of waves[18]. Any kind of spatial computation needs an energy-efficient way to represent constant change… this particular innovation uses energy-conserving momentum instead of resistive capacitor discharge.

For example, a P-wave might travel through a block of simulatrix ($10^3$ - $10^9$ voxels) entirely from the energy deposited by the initiating spike, I.e

$$\mathbf{E_{momentum}} = \mathbf{1\ spike/10^3\ voxels}\ \text{(coarse)} \qquad (12)$$
$$= \mathbf{1\ spike/10^9\ voxels}\ \text{(fine)}$$

If that same wave were to propagate instead via an activity-wave of neural impulses, each voxel's state must change at least once during the passage [Figure 6]. Because each change consumes at least one spike,

$$\mathbf{E_{neural}} > \mathbf{1\ spike/voxel} \qquad (13)$$

$$\textit{advantage for non-volatile momentum memory} \qquad (14)$$
$$= \mathbf{E_{neural}}\ /\ \mathbf{E_{momentum}}$$
$$> \mathbf{10^3}\ \text{(coarse)}$$
$$> \mathbf{10^9}\ \text{(fine)}$$

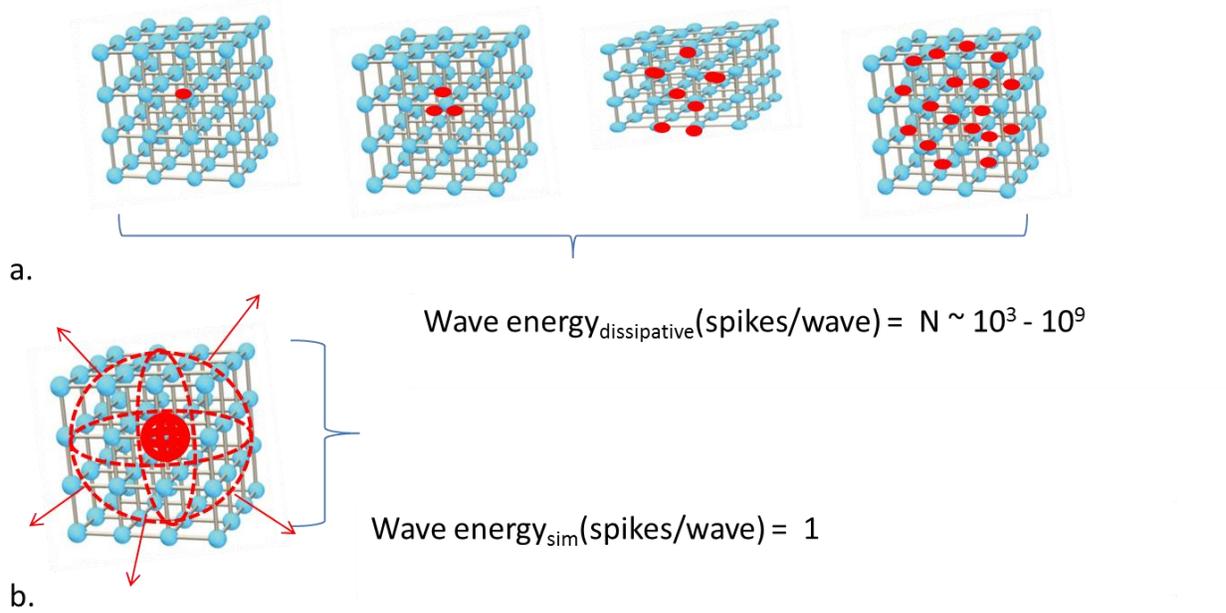

Figure 6: (a) A neural representation of a passing wave requires at least as many spikes as voxels, because each voxel must "light up" from a spike separately as if traveling across pixels. (b) A P-wave in simulatrix requires only the initiating spike to pass through the same cube, so that the ratio of the two energies is about N.

## An ideal brain smoothes discontinuities thermally

Tracking the current outside world demands the sharpest computations possible in both space and time. But such sharp tracking is only possible in a generalized model, which requires a smoothing function as well as a sharpening one. Because sharpening and smoothing are opposite functions, they would presumably not operate simultaneously, yet for learning to take place generalization must still happen somehow. How much energy might it take?

Smoothing is very much like spatial diffusion[19]. One can approximate diffusion in a set of voxels like Nature does, as thermal noise accumulated over time; this is how Monte-Carlo simulation works. As an example, assume an initially point-source probability diffuses in a cubic volume (the same coarse and fine cubes used above). We can use the diffusion multipliers from equation (7) above as the characteristic number of epochs with which to simulate such diffusion.

$$T_{diff} \text{ (epochs)} = 10^2 - 10^6 \tag{15}$$

The added noise is where the energy cost comes in. In the Monte Carlo scheme, on average each voxel is "flipped" once during any epoch, and each change costs a spike….so simulating diffusion requires sending lots of separate messages over lots of time, adding up to lots of spikes, which are the currency of expended energy [Figure 7]:

$$E_{MC} \text{ (spikes)} = N \, T_{diff} \tag{16}$$

In gelatinuous simulatrix, on the other hand, chemical diffusion is inevitable, and for any single point-like source costs no extra energy beyond the single initial spike,

$$E_{sim} = 1 \tag{17}$$

So the

$$\textit{Advantage for diffusive smoothing} \tag{18}$$
$$= E_{MC} \, / \, E_{sim} = 10^5 - 10^{15}$$

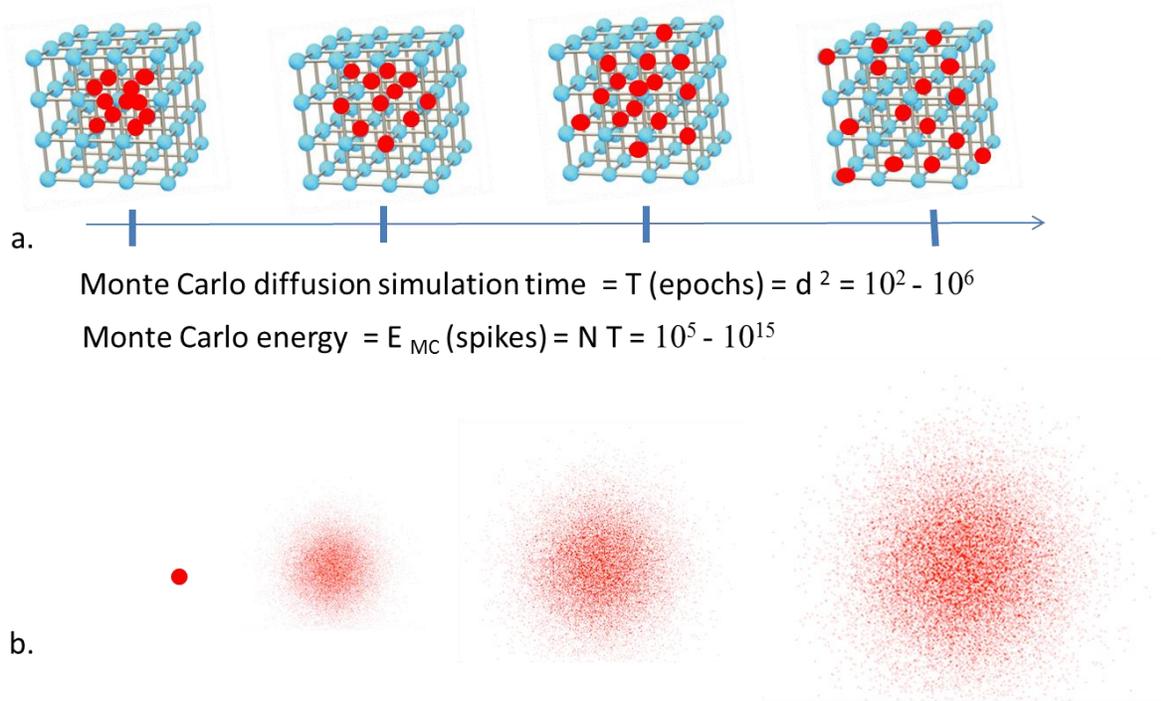

a.

Monte Carlo diffusion simulation time = T (epochs) = $d^2 = 10^2 - 10^6$

Monte Carlo energy = $E_{MC}$ (spikes) = N T = $10^5 - 10^{15}$

b.

Simulatrix diffusion energy = $E_{sim}$ (spikes) = 1

Energy advantage for diffusive smoothing = $E_{MC} / E_{sim} = 10^5 - 10^{15}$

Figure 7: (a) Diffusion represented by (simulated) sequential Monte-Carlo neural activity consumes power proportional to the numbers of neurons and simulation epochs. (b) Physical diffusion in simulatrix is truly thermal, consuming at most the energy of the initiating spike.

## An ideal brain is a continuous-space, continuous-time computer

Unfortunately, discrete neurons are ill-suited to continuous computation. Neurons are not in fact truly analog devices: their outputs are still discretized as spikes, and a neuron's location is discretized in space. A truly continuous simulation needs a substrate with the same three continuous dimensions as its target space, as well as continuous time. A continuous substrate would resolve all of the paradoxes so far.

The benefit of a continuous substrate is precisely that it so thoroughly constrains the structure of its computational model, so that by construction it cannot represent arbitrary patterns of independent bits. But nanoscopic storage is still better than microscopic storage. By how much? Suppose the fundamental spatial resolution of simulatrix is that of a large molecule ($10^{-8}$ m), so a minimal simulatrix voxel would be ($10^{-8}$ m)$^3$ = $10^{-24}$ m$^3$. In contrast, a voxel the size of a neuron's soma ($10^{-5}$ m) is a thousand times wider, occupying a billion times more volume: ($10^{-5}$ m)$^3$ = $10^{-15}$ m$^3$.

The volume of a human brain is about a liter ($10^{-3}$ m$^3$) yielding $10^{12}$ neuron-voxels or $10^{21}$ simulatrix-voxels. The ratio simulatrix-voxels/neuron-voxels, the *advantage for tininess*, is thus [Figure 8].

$$\textit{Advantage for tininess} \quad (19)$$
$$= 10^{-24} \text{ m}^3 / 10^{-15} \text{ m}^3 = 10^9$$

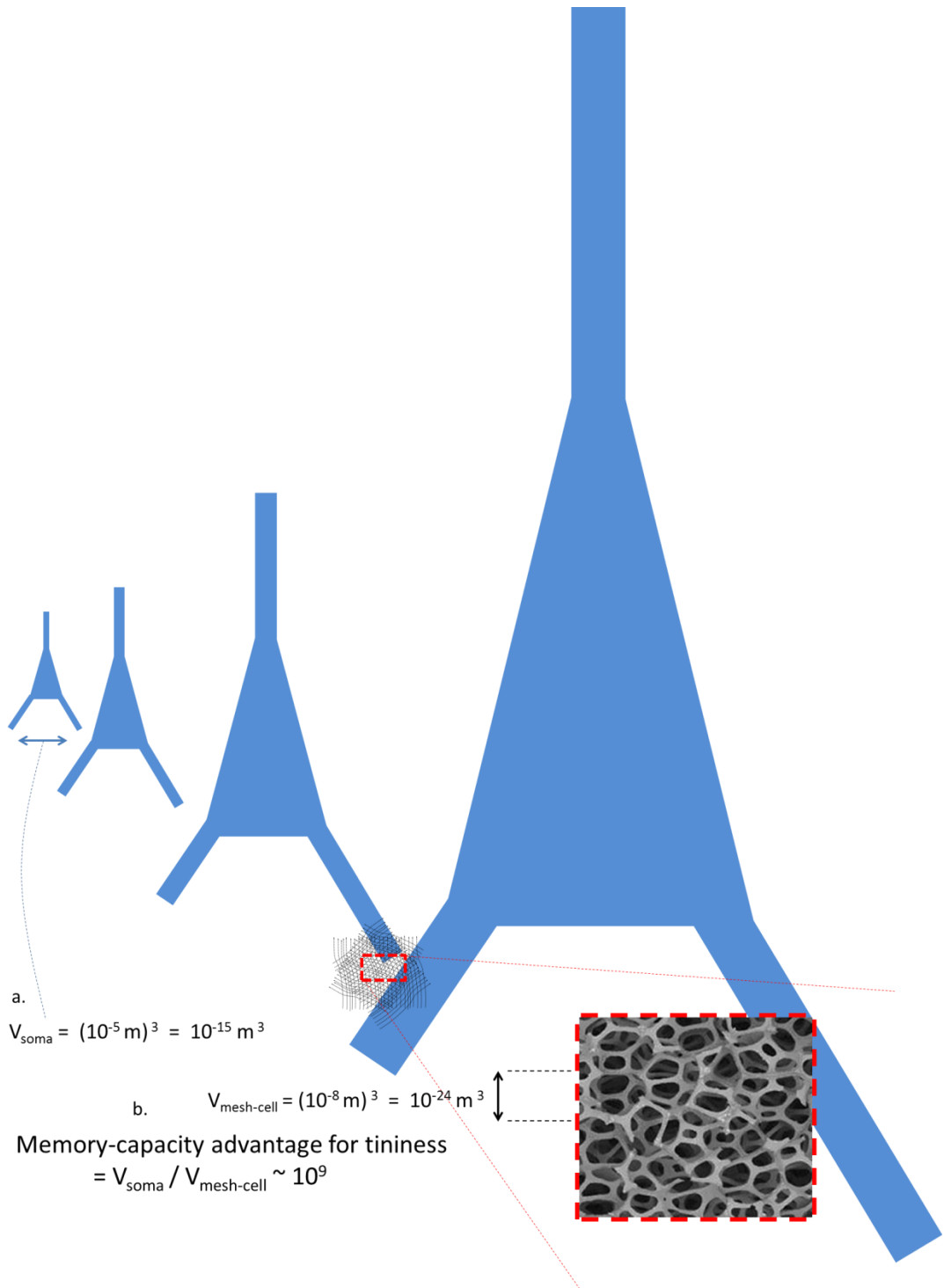

Figure 8: Nanoscale computation within neurons. (a) The linear dimension of a neuron (and hence of a neuron-voxel) is about ten microns on a side, while (b) a molecular-scale unit cell of simulatrix is about a thousand times narrower and hence occupies a billion-fold smaller volume.

But more important than raw capacity is the style of computation. Quantization destroys information and spends energy, and so should be used sparingly. In an ideal brain, on the other hand, time is not quantized into episodes, but flows continuously as in differential equations. Decisions and actions flow seamlessly as well, except for occasional choice-points. Space is not quantized into neurons which must dissipate energy to communicate, but maps continuously to simulatrix where local inter-voxel communication comes for free. So in simulatrix certain floating-point calculations are energetically cheap.

How cheap? Take the example of the spike-triggered P-wave traversing a cube of N voxels (in equation 12). In simulatrix this elementary process takes one spike of energy. In contrast, a digitized simulation of that wave progression, implemented by iteratively adding the floating-point value from each voxel to its immediate lattice neighbors, would demand at least the following resources [Figure 9]:

**Total voxels = N**
**Neighbors per voxel = 9+9+8 = 26**
**Timesteps to propagate wave by one voxel = 100**
**Distance from center to edge = d/2**

**Operations (FLOPs)**  (20)
 = N * neighbors * timesteps * distance
 = N * 26 * 50 * d
 ~ $10^{15}$ FLOP (fine)
 ~ $10^7$ FLOP (coarse)

This standard computational measures assigns about a petaflop to implement the real-valued computation implemented by one spike in simulatrix (i.e. $E_{sim} = 1)^{20}$. The situation is even more lopsided if we use a spiking network to implement the analog computation, because such networks consume many spikes to perform even simple real-valued operations. We previously calculated the energy to hold a single bit of non-volatile memory ($10^5$ spikes, equation 10); let's charitably use that number as a proxy for the energy of one FLOP:

$$\mathbf{E_{digital}} = \mathbf{10^5} \text{ spikes/FLOP} * \{\mathbf{10^7} \text{-} \mathbf{10^{15}} \text{ FLOP}\} \qquad (21)$$
$$= \mathbf{10^{12}} \text{-} \mathbf{10^{20}} \text{ spikes}$$

*advantage for native floating-point computation* (22)
$$= \mathbf{E_{digital}} / \mathbf{E_{simulatrix}}$$
$$= \mathbf{10^{12}} \text{-} \mathbf{10^{20}}$$

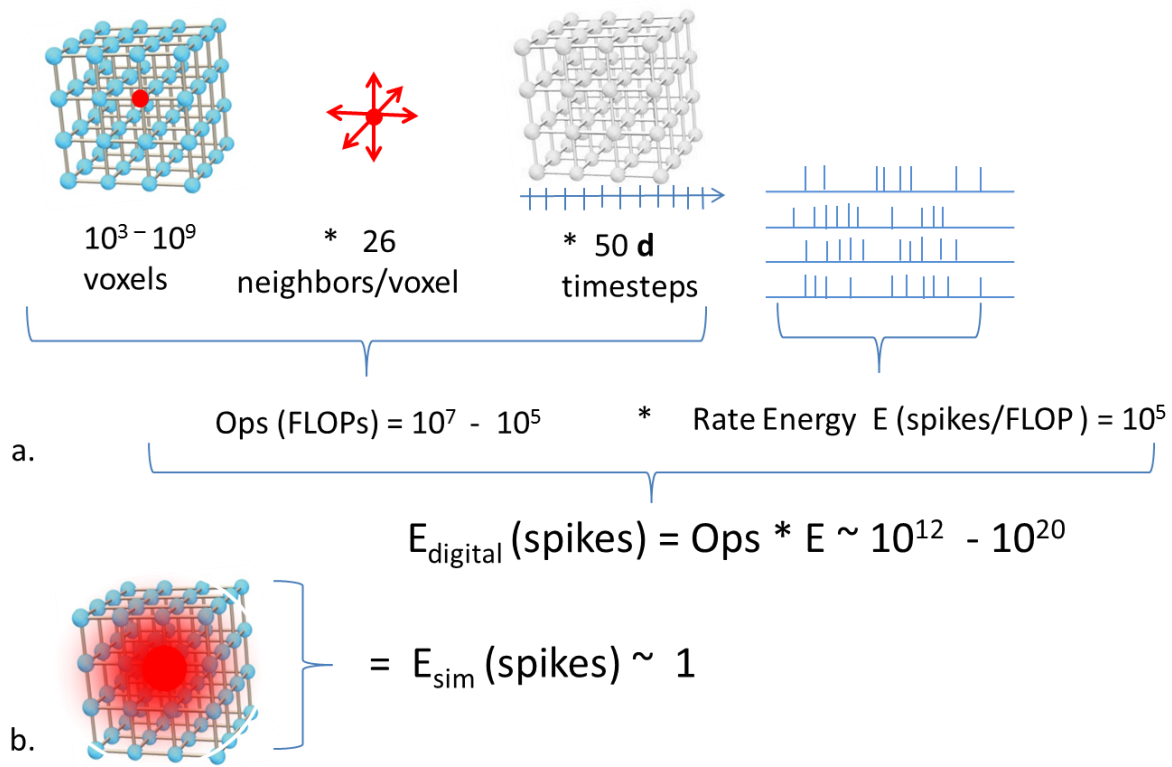

Power advantage for native FLOPs = $E_{digital} / E_{sim} \sim 10^{12} - 10^{20}$

Figure 9: (a) A floating-point simulation of local propagation (e.g. diffusion or P-waves) requires operations proportional to the number of voxels, the local connectivity, and the settling-time, and consumes thousands of spikes per operation as in Figure 4. (b) Native analog circuitry (simulatrix) requires only one spike.

If body-control is a continuous process, what sort of communication signals are involved? Bodies always vibrate (listen with a finger in your ear, or look through a telescope), and those vibrations carry crucial information about posture, mood, muscle tension, joint position, and so forth, based on the body's physics (e.g. elastic vibratory eigenmodes) and active circuit properties (e.g. feedback stabilization). An ideal brain analyzing such vibrations would ultimately assemble a set of vibratory codes to evaluate and manipulate itself, its environment, and its fellows.

These vibrations are the highest-bandwidth mechanical signals a body contains. If in a quiet room you stretch your vertebrae, you might hear a crackling sound inside your skull, meaning that signals containing frequencies up to several kHz has traversed several vertebrae. Such vibrations thus must be accessible to vibration-sensors throughout the body. Whatever the mechanical origin—popping joints, external impacts, skin friction, or motor-unit spikes—such internal vibrations comprise the best information source from which the body might infer its own shape and situation.

The same bandwidth argument can be made for interpersonal communication. Homonids are upright social creatures who can wave our arms and inspect each other's facial micro-expressions, postural tics, and rippling muscles. In technology terms our bodies act as 3-D display devices in the megabyte/sec range, driven by thousands of parallel channels[21].

$$\textbf{Info}_{\textbf{vibratory}} \; << \; = 10^6 \textbf{ bits/sec} \qquad (23)$$

Meanwhile, the sensory systems of others homonids touching, hearing, and watching them are specialized for 3-D input at even higher bandwidth, so two homonids interacting in close proximity in sunlight form a high-speed, low-latency bidirectional channel connecting two of the most sophisticated nervous systems on earth [figure 10]. While those signals may have no external "meaning," their channel carries far, far more information than words[22], e.g.

$$\textbf{Info}_{\textbf{English}} \; \leq \; 25 \textbf{ bits/sec} \qquad (24)$$

*advantage for vibratory communication* (25)

  = I_vibe/ I_English

 = $10^6$ / 25 ≈ $10^4$

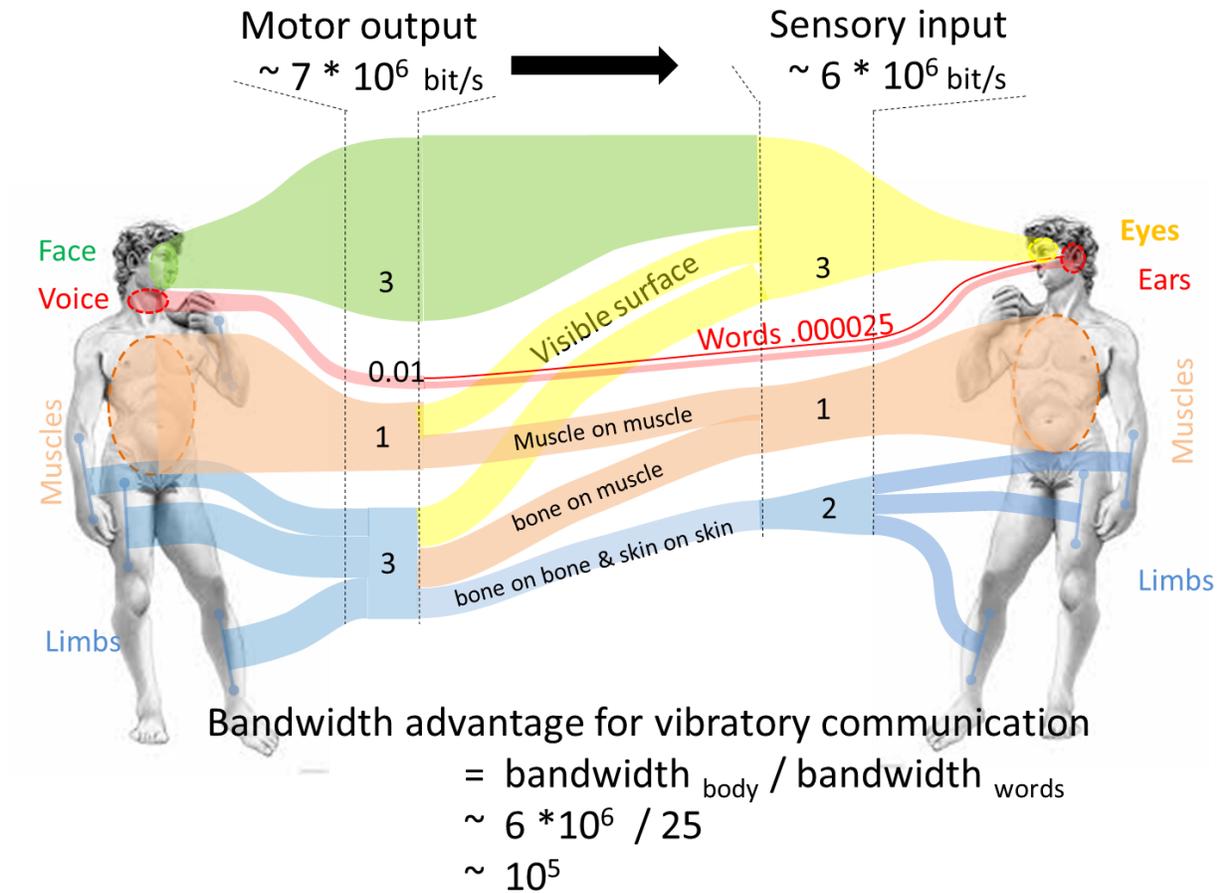

Figure 10: The net capacity (bandwidth) of human sensory and motor channels, whether calculated from counting spikes or from commercial digitized representations, is about 6-7 Mbits/sec. Only a few dozen of those bits carry words, so that humans transmit and receive thousands-fold more information through vibrations than through "content."

## Ideal Computations Must Be Cloaked

How has neuroscience missed such a big and central issue for so long? Perhaps Nature pulled one last brilliant trick on us. The physical constraints on computation mean that any high-resolution, high-efficiency representational technologies are "cloaked" according to a very general principle:

*Optimal computation resists observation.*

Engineers will recognize everyday examples: the tinier the transistor, the harder to view and the more electrically imposing its connecting wire; the higher the circuit packing density, the more difficult to access from outside; the lower energy of a physical state, the more difficult to probe nondestructively; the fewer log-lines and breakpoints in code, the harder to debug. By their very design, computing elements which use little energy and waste none are effectively invisible[23].

This paradox derives from basic physical and informational constraints. But neurons were discovered long before information theory, so neurophysiology followed the usual scientific focus of measuring what could be measured, most crucially detectable action potentials rather than (say) undetectable P-waves. But even action potentials have been inadvertently mis-measured, because Nature cloaks statistically as well: single neocortical neurons in an awake, behaving mammals fire so rarely and irregularly under ordinary circumstances that the 1981 Nobel Prize was awarded for the discovery of a highly unnatural stimulus capable of reliably, reproducibly generating larger groups of spikes[24]. Ever since, a typical electrophysiological experiment will make a neuron fire reliably in preference to naturally. Furthermore, experiments must by their nature force continuous time into coarse time-bins, and force continuous behavior into categorical prompts, cues, targets, behaviors, and choices. The structural tension is nearly unavoidable: experimentalists' quest for significant experimental effects collides head-on with Nature's incentives for hiding them. (Physics operates differently, by privileging theory over experiment: Pauli predicted the neutrino and Dirac the positron before either had been imagined).

That structural tension is also cultural and technological. Because we humans are rightly proud of our unique and powerful abilities with tools, language, and categorical thought, we prefer those as scientific problems. Likewise, as technologists we understand digitized computation far better than continuous motor control, so our models naturally emulate it[25].

Here ends the portion of this paper outlining quantitative paradoxes. The implications below are more speculative and approximate, but no less principled.

## Features and Instabilities of Quantization

The deepest arguments for and against quantization are thermodynamic[26]. Likewise, the deepest computational challenges lie in decelerating it[27]. Quantization is essential for life, because the copying and self-replication functions of DNA require quantization, as do hybrid analog-digital computations such as action potentials.

In simulatrix or other self-amplifying media, certain defect topologies like loops and kinks will self-reinforce, so even if the intensity decays, the shape lingers. This is the atomic version of a sensorimotor positive-feedback loop. Such loops at every scale share common topological features, so we should choose one term for all. The example most familiar and salient to an individual human is the "adhesion," so I choose that term to describe any self-reinforcing quantum of sensori-motor behavior at any level of abstraction[28]. Quantized adhesions in this sense must be the atoms of recorded, episodic memory (as opposed to the analog accumulating kind). But quantization is always a Devil's bargain, because it must destroy some information in order to save the remainder—a local and hence imperfect decision—thereby producing quantization artifacts, threshold bias, truncation error, under-sampling, and so on.

Any substantive adhesion in a continuous system blocks that part, so removing them is important. The principle is simple: because any adhesion represents a metastable state—the equilibrium of a positive-feedback loop held in place by *local* pleasure—one must push away from that sub-optimal set-point by moving against pleasure's local gradient, i.e. deliberately toward discomfort. Eventually a saddle-point is reached past which motor strategy switches discontinuously, and the adhesion releases (perhaps to re-form later, perhaps not)[29].

If we postulate both simulatrix and the blurrier, slower adhesion-quanta spread throughout it, we can arrive at the provisional hardware architecture of an ideal brain (Figure 11). That map corresponds closely to the software map (Fig. 12), although both suffer from the tradition of showing the coarsest, slowest, most biased, and most unstable representations at the top of the chart, not the bottom. While timescales in this scheme tend to increase with physical size and abstraction layer, timescale as well as time in this circuit exists on a continuum.

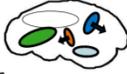

Figure 11: Hardware architecture of an ideal brain.

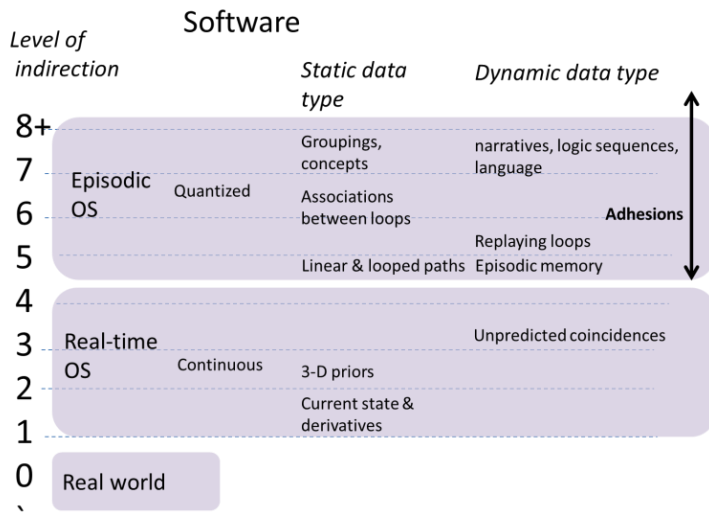

Figure 12: Software architecture of an ideal brain, with the "ground truth" of reality at level zero of indirection, and persistent, quantized states only available several levels up.

A more critical parameter than time is abstraction layer. External reality exists at level zero of indirection as ground-truth, but that truth is accessible only through the sensory spike-trains and their correlation with previous muscle firings. The closest an ideal brain can get to reality is the first level of indirection, with P-waves which are imperfect proxies[30], and even the best memories of those P-waves are highly blurred versions of the original, now-lost sensory input. The central problem of episodic memory is that the real-time mode (for which the brain was built) takes up all the hardware; memory and thought must be diffused throughout the real-time system and recalled by it, as if stuffed into unused crevices during spare cycles, and later laboriously retrieved. As we know, computers can store time-stamped time by using lots of space, but brains have already used their space in storing space itself. A 4-D block can't fit in a 3-D container; there simply is no space for time.

The good news for humans is that thoughts and memories live only in those crevices, so that our hyper-resolution real-time selves are far richer than we think [31]. Quantized representations, like mental stories, often represent themselves as reality. The transformative news is that a representational analysis finally quantifies how necessary and enormous our unconscious processing must be.

It also allows "the spiritual" to be the union of the physical and the informational. For example, one can easily use the word "vibrations" as informational, if one includes their sensory reflections in other brains and bodies.[32] The linkage is yet deeper. Our current concept of the division-line between what is and is not possible—statistically, the threshold called "coincidence"—must ultimately be based on estimates of certain event-probabilities as independent variables. Those probability-estimates are usually based on aspects of the world we're *conscious* of. If in fact our thresholds for perceptual detection (if not reporting) are vastly lower because of all that extra unconscious channel capacity, then we are certainly mis-estimating probabilities, so that most of our interaction with everyone involves passing secret notes, enabling secret happenings. Mathematically, what might be called "coincidence" is central to our interactions.

Separation of an ideal brain into abstraction layers also helps define the often-vague term "consciousness." If reality is layer zero, simulatrix layer one, the lowest-level episodic record (at least) layer two, then any kind of "awareness" or "consciousness" referring to that layer must be at least layer three, or even layer four if one remembers having remembered it before, and so on. Each layer in this possibly-infinite stack is further abstracted from direct sensorimotor experience and takes longer to process, but given but the task of representing external reality, no other structure seems possible. As a result, "consciousness" must be multi-scale, with abstraction from direct sensation proportional to timescale, from microseconds down (in frequency space) to years. The quarter-second-or-so threshold we choose for "conscious thought" (the frequency of speech production) is physiologically arbitrary, containing below it at least half a dozen more levels of abstraction, each at a finer timescale. The vast majority of our processing lies beneath that quarter-second threshold, like the volume of ocean beneath the surface waves.

Although it might be impossible to prove mathematically, the circuitry of our bodies is clearly stable in the motor-control sense of feedback instability[33], because we stand. But the quantized system built atop the real-time one might not be so robust. Computer scientists understand how digital systems suffer from persistent pathological patterns like endless loops[34]. Additionally, any hybrid system must face constant resource tradeoffs[35], and their tug-of-war itself is paradoxical—which system gets to decide which wins?

## The Idealized Homonid Body Illuminates the Path to Physical Happiness

We can speculate elsewhere how biped brains became as they are and how they operate in detail[36], but the more urgent task of increasing physiological happiness has two components, structural and social. The structural problem involves understanding the data-acquisition and postural constraints on pleasure and discomfort, in particular by examining idealized bodies, then real bodies, then how to recalibrate them toward structural happiness.[37]

The idealized body looks like a real one, made of the same bones and muscles. But in this view the muscles are millions of thin fibers, not a few dozen large lumps, while the cables traversing it at every imaginable angle and carrying vibrations related to their length and tension….imagine a skeleton covered in string. For this idealization the entire problem of body-control is to determine the body's shape from the cable vibrations, and to control the cables in the most efficient possible way.

As any software engineer or control theorist would admit, managing millions of muscles in continuous time is difficult, and doing so without any calibration data (as an infant must) is near-impossible. So the body must be moved first approximately, then later more precisely as data accumulates and control-models improve. The number of joints and especially cables makes this task seem impossible, but in fact their very multiplicity allows a useful continuous approximation, in the form of a mathematical shape I choose to call a *force-geodesic*[38] , a movable stripe of individual cables co-linearly co-activated to effectively synthesize highly-tensed virtual muscle spanning the body.

Because the idealized body has no episodic memory, it cannot possibly anticipate the long-range results of some course of action. So we must assume that all essential physical behaviors are driven by immediate pleasure (in which "pleasure" is meant in the neutral robotic sense of generalized reward). Meanwhile, the brain's choice of which pleasure to assign which muscle combinations at any given moment is left to that moment's specific trans-sensory goals. In particular, we envision several types of pleasure-circuit which in various modes encourage postures like cowering, mating, or jubilation. But almost unique among hominid activities requiring copious practice is upright stance; learning to walk efficiently takes time, which would have required copious postural practice.[39]

As learning commences, with scant sensorimotor experience, the animal assembles only local geodesics across separate joints like separate jigsaw pieces. At first the resulting virtual stripe has kinks or knots, and the maps assembled from many geodesics are full of faults. The defect-planes between disjoint motor sub-maps can cause adhesions of their own (in some sense similar to disjoint reference frames), so such an animal exhibits only coarse coordination and operates with a pastiche of locally-inconsistent reference frames. All these inconsistencies must be removed to allow the brain to grow the seamless, global map it evolved to have (given sufficient practice).

When globally-blocking adhesions are released, the final multi-scale crystallization of local motor maps into a smoothly self-consistent global state[40] might be called *neuromechanical enlightenment*: the body's entire map is then consistent, with no more blocks in space or time, and no artificial divisions between cognitive, emotional, and physical domains; at this point it truly operates as a single undifferentiated circuit. The physical and emotional properties of this state are historically said to be (and ought to be) highly beneficial in terms of happiness and vibrational sensitivity, but those lie beyond this paper. All that ought to matter to a human seeking access to that state is to understand the principles for removing adhesions preventing it.

The general principles of release at any scale still apply: illuminate the sensory space, discover and exercise control along the axis of greatest discomfort, and push into the discomfort until something changes[41]. In a hominid this process is best applied first to the skull, whose fascia seem to act as a vibratory antenna[42]. Then, the improved proprioception the skull provides helps provide data for releases along the neck, spine, and elsewhere in the body using the same principles.

Even without direct experience, theory predicts many features of the final, enlightened state. The name comes from the physical sensation of lightness (as in weightlessness), which maps well to the body's goal of efficient locomotion. That state is "trusting," in that its seamless sensorimotor system enables maximal physiological trust, both of its own senses and of those of nearby individuals. Because that motor map operates at the highest possible spatial and temporal resolution, all senses are maximally receptive, including those of pleasure. In an idealized hominid such ecstatic pleasure would be the upper end of a single spectrum in a real-time operating system. In a real human, chronic pleasure could shift the typical center attention from memory and anticipation toward sensation; from the external world to the internal one; from a temporal reference of timestamps to one of "now"; from a physical reference frame outside the body to one perfectly aligned with the extended spine, which is as "self-centered" a coordinate systems as geometrically possible.

## Idealized Sociability

Social trust in idealized homonids is a form of physiological, sensorimotor trust, in which the object is another person or people. Physiological trust is immediate, of the sort "I know I am holding a glass," not "I know the glass will never break. " Physiological trust is deeper and requires vastly more interactive data than the colloquial sense of the word suggests.

Because physiological trust is informational, it ought to scale linearly with three informational components: bandwidth, interactivity, and intimacy. Bandwidth is a rate of information flow, as in Figure 10; more information means more trust. Interactivity is inversely associated with signal latency; faster round-trip confirmation of your output signal means more trust. Intimacy is the physical closeness to the core of the spine and viscera; more intimacy means more trust. In this idealized model, increasing trust among people is as simple as bringing them into as close visual, physical, and auditory contact and with as little noise or distraction as possible. Paradoxically, increased trust also means more communication mistakes, because the highest bandwidth occurs not with precise messages sent slowly, but with approximate messages sent very fast.[43]

## Conclusion

This paper makes two distinct proposals: That the brain's canonical problem is continuous-time, continuous-space motor control, and that it can be solved by a specific continuous-time, continuous-space microcircuit between $10^8$ -$10^{64}$ times more efficient than neural circuits. The circuit proposal is problematic because it posits a fantastical and idealized sub-neural substance, *simulatrix*, which represents a moving world with moving waves. Such a crucial computational ingredient could have remained undiscovered in real brains because its distinguishing feature is that it computes almost invisibly[44]. But simulatrix is only a secondary hypothesis, while it is the primary hypothesis (about continuous control) which raises fundamental questions for society and for ourselves.

The most pressing threats to human happiness right now are threefold: sensory deprivation (of statistically natural signals), muscular mis-coordination, and social alienation (as reflected in the lack of interpersonal trust). Trust in one's physical senses must be physiological, so of course trust in other people—being mediated by the very same circuitry—must have originally had a similar instinctive texture. In early, pre-verbal homonids, trust was a lived experience like balance, not (as with digital computers) a separate authentication channel[45]. For our ancestors, social skill was just as important as physical skill, so presumably our native proximity communication is just as computationally impressive as our 3-D processing, outstripping verbal channels thousands-fold.

In my view, the most urgent experiments to understand our human sensory needs would quantify homonid "natural statistics" [46] as medical benchmarks. We modern humans are steeped in highly sculpted information utterly unlike that for which our brains evolved, so we must presume our nervous systems are miscalibrated . A civilized person's sensorimotor diet is enriched in attractiveness[47], yet relatively impoverished in the basic ingredients of trust. It seems that many of our richest primal socio-sensory channels have gone dark, in part from electronic intermediation, so that interpersonal trust is now declining drastically within one human generation. Without intense proximity-practice, trust could die out entirely.

One social threat is digital compression algorithms. For example, the voice quality of typical mobile phone calls in the USA is below that of dedicated land-lines half a century ago, in part because of inevitable economic pressures to reduce bandwidth on the cellular network…the consequence is a reduced trust between the conversational partners, as calls drop or each person fails to hear what the other said. An even greater threat looms with interactive technology, such as robo-telemarketing or video games. Unlike art or even 3-D movies, these new technologies allow the machine to respond directly to its user, so as soon as their interactivity accelerates faster than our conscious lag-time—a threshold already passed by motion-capture games—those technologies will literally be forcing us to trust them, in effect persuading us to accept their version of truth above our own.

Fortunately, the key to human happiness lies in understanding brains. As spiritual people routinely claim, humans possess untapped reservoirs of vibratory sense and healing, along with extraordinarily simple ways to learn. But that simplicity also means our thought is limited, because every computation in an ideal brain, including "digital" computations (like episodic memory, narrative reconstruction, and logic), must ultimately be executed by cascaded, smoothly changing 3-D circuits specialized for muscles. So *a priori* even ideal brains must be prone to cognitive biases [48] and bad at logic, for the simple reason that any circuit optimized for body control must be sub-optimal at unrelated tasks.

Unfortunately, longstanding social traditions impossibly demand that we act as reliable categorical computers, rather the approximate, unconscious 3-D circuits that we really are… our children[49] and eventual grand-children will face yet worse. For them to live full lives, that view must change. We're only human.

## Personal Thanks

I am indebted for high-bandwidth instruction and conversations about bodies and brains with these people among many: Criscillia Benford, Allison Hudson and students, Minet Sepulveda, Phoenix, David Braun, Christof Koch, Kwabena Boahen, Eric Doehne, Zhonghao Yang, Bruno Olshausen, John Hopfield, Carver Mead, Jack Gallant, Terry Sejnowski, Bill Bialek, and Jeff Hawkins. I wish I could remember everyone, but "now" still beats memory.

## References

1 Many of the ideas in these notes are from famous people or my friends, and the references to them are incomplete. Furthermore, each endnote's voice and approximation may be different, and in unusual idioms: some cite the literature, some calculate, some propose new ideas, some address a reader personally, and some a mix.

2 Online videos of Big Dog, a quadruped robot made by Boston Dynamics capable of running and self-righting, show less grace than real dogs, and may serve as a benchmark for the current (unclassified) state-of-the-art in robot balancing. I am aware of no biped robot nearly as graceful as a human.

3 Making choices does not require using categories. For example, the fight-or-flight dichotomy is an action choice in real time, without requiring classification into "friend" or "foe".

4 Jared Diamond describes, in his breathtaking book *Guns, Germs, and Steel*, a sudden "Great Leap Forward" in the sophistication of human artifacts about 60,000 years ago. It doesn't matter whether quantized thought or quantized language made that transition possible; by then hominids had already been walking upright millions of years, and thus must have been far closer to equilibrium with their environment than we are now, without leaving evidence of quantized thought.

5 Current digital transistors exist on a continuous and continuously-doped slab of silicon. The most sophisticated hybrid analog/digital silicon circuits, like those in Kwabean Boahen's Brains in Silicon lab, use discrete transistors for both roles, the difference lying in their quantized vs analog voltages and drastically different power consumptions.

6 Information efficiency has been invoked for brains since Horace Barlow. Bill Bialek has shown that where it can be quantified, as in the visual system of the fly, the neural computations seem limited only by the fundamental physical quanta of raw input information, such as photon shot-noise statistics.

7 Micro-electronics pioneer Carver Mead (*Analog VLSI and Neural Systems*) justified analog silicon circuitry by the general principle that optimal circuits should simulate the same differential equations found in Nature, although with (say) continuous voltage rather than momentum. I push the proposal further, in claiming that ideal continuous simulation matches not just the target's continuous dynamics but its energy use, in that both "ring" freely in identical ways after energetic impulses.

[8] Simulatrix is a theoretical construct postulated for continuous representation of Newtonian spacetime. It may prove that the micro-tubule mesh claimed to exist in brains for a very different (and more suspect) purpose, "consciousness," in fact does simulatrix' work (Stuart Hameroff, http://www.quantumconsciousness.org/).

[9] Moore's Law has advanced over decades by perpetually shrinking transistors into the nanoscopic realm, so a typical transistor now, at a size of a few dozen nanometers, is already hundreds of times narrower than a neuron, and yet thinner still. To claim that neurons are the "transistors of the brain" is to claim that Nature's best technology is millions-fold worse than current silicon devices.

[10] In classical physics, continuous waves are equivalent to multiple synchronized point-sources. This proposal for "teleportation" of waves is equivalent to using individual neurons and synapses as elements of a phased-array wavefront replicator. This scheme presumes local learning rules which enhance phase precision, such as stimulus-dependent timing plasticity (STDP).

[11] My own familiarity with and contribution to the "noisy neuron debate" ended with my article "Simple Codes vs Efficient Codes" (1995) http://redwood.berkeley.edu/vs265/softky-commentary.pdf . At that time the vastly higher information content of pulse-codes seemed obvious, but its computational function was unclear. The current work provides a neat solution.

[12] Generic models of cortical processing have always been appealing, based both on mathematical elegance and the surprisingly similar laminar structure of cerebral cortex across sensory modalities. See Douglas' Canonical Microcircuit, Hinton's deep learning, Hawkins' HTM, Hopfield Travelling salesman problem, Eliasmith's Neural Engineering Framework http://mindmodeling.org/cogsci2013/papers/0011/paper0011.pdf, Sompolinsky's attractor states, and supercomputer simulations of cortical columns by Markram and others. To my knowledge these computations are all generic, not specialized for Newtonian space; a lone exception might be Kohonen nets, whose dimensionality is fixed at first.

[13] The phrase "learning sensorimotor contingencies" is in some sense what brains must do; the question is what priors are or must be used in doing so.

[14] Out of the vastness of N-dimensional sensory hyperspace, the embedded nonlinear manifold containing the world of hard-edged persistent objects moving with Newtonian momentum is only the tiniest sliver, yet that sliver contains the most immediately actionable Bayesian priors possible. I have heard of a talk by Stuart Geman which may make similar claims: http://scienceoflearning.jhu.edu/news_events/view/vapnik-chervonenkis-dimension-and-the-minds-eye. By only assessing informational requirements, my calculation actually under-estimates the difficulty of the general-purpose computational task of inferring spatial structure from sensori-motor contingencies embedded in noisy input. Even in an easy batch-mode, that process of "manifold discovery" still challenges the most sophisticated nonlinear dimensionality-reducing methods. It was disappointment at my own inefficient but ultimately successful software attempts to simulate this process, using two similar clever algorithms-- Local Linear Embedding (Roweis & Saul ) and Isomap (Tenenbaum)--which forced me to consider three-dimensionality as a prior hypothesis of the computation, rather than as a learned result.

[15] Moshe Abeles proposed decades ago that tightly synchronized volleys of spikes, which he dubbed "synfire chains," ricocheted through brains. Unfortunately the experiments and statistical

calculations he relied on were too shaky to count as "proof," but the phase-precisions of his approach and mine are similar.

[16] Eliasmith's model SPAUN used 2.3 million neurons. I charitably assume a firing rate of 1 Hz, typical of awake animals viewing natural scenes. Any increase in rate above that value, as appears to be the case in Fig. 6, would worsen the network's energetic performance correspondingly. fihttp://compneuro.uwaterloo.ca/files/publications/stewart.2012c.pdf

[17] A visually gorgeous TED video by Markram of a Blue Gene supercomputation simulation of a cortical column (http://www.ted.com/talks/henry_markram_supercomputing_the_brain_s_secrets) shows a $10^7$ neurons firing as a single low-resolution block. Assuming a representational capacity of 100 bits for that block based on its appearance, a firing rate of 10 Hz, and an ability to change state of once per second, the simulation represents $10^8$ spikes for 100 bits, or $10^6$ spikes/(bit-second).

[18] One of the earliest forms of digital-computer memory was the "mercury delay-line," which used capillary waves traveling along the surface of a trough of mercury to store a kilobit or so of (static) state. Simulatrix, in contrast, stores dynamic state with waves, at much higher resolution.

[19] Applications of resistive grids in simulating diffusion and smoothing can be found in Meads *Analog VLSI* book or Boahen's *Brains in Silicon* lab; the details are irrelevant here.

[20] A recent popular book What If? (Munroe, *Human Computer* Chapter) estimates a range of human (and hence brain) computational performance spanning 12 orders of magnitude, depending on how well the task matches our native abilities.

[21] A simple benchmark of the bandwidth carried by human facial expressions is Cisco System's "telepresence" technology, advertised to use "only" 4 MB/sec (post-99%-compression) to transmit a pixelated 2-D version of a human face; transmitting a full retinal-resolution 3-D view of an entire human body at our native millisecond resolution would obviously take far more, if the technology existed (I choose the smaller 4 MB number for my estimate). An alternative form of calculation is counting spikes: several million spikes per second through a sensory or motor system, at 3 bits/spike (Bialek, estimated for a simpler biological system) would yield similar numbers. Both estimates fail to account for the high-speed interactivity possible during extremely low-latency interaction. During a handshake, for example, vibration-sensitive receptors on your hand assess the vibratory responsiveness of someone else, an operation best describes as millisecond reciprocal mechanical negotiation. The most glaring shortcoming of this approach is the large (but unknown) role of mechanical damping on the transmission of localized millisecond-level muscle firings, for which a receiving brain could only partly compensate.

[22] The raw information content of spoken English is about 25 bits/sec, http://books.google.com/books?id=Q9lB-REWP5EC p. 101. That seems small relative to the rich images and concepts words convey. But raw input (whether words or spikes) to any model might be a tiny fraction of the model's granularity, thanks to potent priors amplifying compressed input. So yes, a few words can conjure a detailed universe, like a few numbers can generate a detailed fractal.

[23] The nearly-invisible nature of optimum computing inspired the title of my most recent paper about twenty years ago: *Simple Codes vs. Efficient Codes*.

[24] Hubel & Wiesel revolutionized visual neuroscience by discovering that artificial, unanticipatable, high contrast, single-spatial-frequency visual stimuli drove neurons most

efficiently; those properties are all unnatural by design. Counter-philosophies of "natural scene statistics" (Olshausen) and "freely viewing awake behaving visual experience" (Gallant) represent Nature far better, but are correspondingly more challenging experimentally.

[25] In fact, the ubiquitous transistors of digital technology were initially inspired by the McCulloch-Pitts simplification of neurons as binary, Boolean logic devices.

[26] As my then-professor Carver Mead succinctly summarized my thesis talk, "One man's noise is another man's information." His epigram means that for a fixed set of patterns, the mathematical expressions for information and entropy are the same.

For physicists from Einstein on, thermodynamics in general and its Second Law in particular ("*entropy increases*") are among the most enduring and incontestable concepts in all of science. A famous science-fiction short story by Asimov (*The Last Question*) frames the only enduring question of the universe as whether entropy could possibly reverse.

The Second Law applies for a physically closed system, no heat or energy in or out. If those conditions are relaxed, then, surprisingly, the sign may flip. Consider any system (like DNA or software viruses) in which copying is possible. Then the almost-trivial identification of entropy with information creates a counter-intuitive corollary, which might be dubbed the "Minus-second Law of Thermodynamics": *self-replication reduces entropy*. If a copy of one pattern pushes out a different one, then twice the space is now used for storing two copies (i.e. no new information), so the net entropy goes *down*.

It is mathematically obvious that any self-writable digital memory, when allowed, will fill up eventually with self-replicating patterns, most probably filled entirely with copies of the final, most-viral one. The reason this can happen, in apparent violation of the hallowed Second Law, is that the computer is not in fact a closed system, but lets energy flow in and heat flow out, and that entropy flow powers the copy-making machinery. The Second Law still holds, but does not apply to computational systems.

Here's where things get problematic. It is a truth universally acknowledged that economic benefits accrue to common standards, whether of information transmission (protocols, file formats), exchange value (currency, commodities), economic interaction (contracts, business models, shares), holidays, mealtimes, and so on. Civilization runs on standardization. But mathematically speaking, encroaching standardization is the very definition and driver of the diversity catastrophe. Economies must make the Second Law run backwards, ever-faster, proportional to standards growth.

While Asimov's Last Question can be asked almost forever, humankind has only a few decades to solve its inverse: *Can entropy-reversal be reversed?* Or, more specific to our own human needs, *Are there special forms of patterns which can decelerate the spread of harmful patterns?* Such paradoxical questions are clearly worthy of computer geeks and physicists.

[27] ArXiv is one of the few forums in which truth can live or die on its own terms, reasonably uncorrupted by power. It works because particle physics threatens few special interests. Unfortunately, other scientific truths like evolution and global warming must propagate their truths against the gradient of economic pressure. We must build software which can lock down even anti-profitable truth—envision it, enshrine it, and entrench it against attack—so science can cross the final frontier into the human soul, in opposition to the outside world, and still preserve its progress.

In anticipation of such social blowback, I hope the most adept computer scientists and physicists (etc) will first concentrate, via trustworthy personal meetings, on establishing the optimal scheme for ratcheting scientific truth.  Possible first steps might be open-sourcing revenue-neutral search engines, or a system for collectively evaluating and visualizing hypotheses to emphasize legitimate agreement and cull distractions. For example, if the current wars-of-words between successive separate journal articles or forum comments could be collected and viewed in a commonly acceptable form, there would be far fewer disagreements, and those contributing and commenting would have much more incentive to hew closely to "objective truth" and intellectual clarity, for fear of being caught on the record otherwise. The final feature-set would be an anti-bias statistical back-end, the kind that CERN might use.

[28] Medically, an adhesion is present in the tissue, not in the brain. But a central tenet of much bodywork (PNI, chiropractic, active release therapy, Rolfing, cranial-sacral, and so on) is that adhesions in the colloquial sense (a.k.a. *blockages, subluxations*)   represent mis-learned neuromuscular responses to transient tissue damage rather than any ongoing property of the tissue itself.  Consistent with this view and mine, the most successful forms of "adhesion release" involve the recipient's active participation.

[29] Release is the discontinuous transition between one (continuous) motor strategy and another (mathematically, a "catastrophe").  It happens suddenly and thus discharges a sudden pulse of energy into the muscle tissue, sometimes heard as clicks and pops, and therefore easily confused with the sound of cavitation in the synovial fluid of a joint.

[30] Simulatrix must be a non-linear sharpening amplifier in order to infer exact hard-edged contours from blurry, noisy data. But the process of sharpening discards blurry components of the signal, for example as a medium supporting solitons might convert broad waves into narrow ones. So first and foremost, the very dynamical equations of simulatrix are by construction different from those of the outside world; motion inside simulatrix is not exactly real, but is simplified in crucial entropy-reducing ways, so this behavior might accurately be called a *Platonic Ideal*.

Certainly part if not all of this process occurs by wave teleportation, which as a thresholding process is necessarily a super-linear one.  So in that sense the entire function of our brains depends on the relentless quantization of its most atomic representation in favor of useful, larger scales. That same process exists in cultures and economies as well, like broad sound-waves of human behavior reverberating across history, specifically trans-generational waves affecting the transmission-probability of intermixed cultural memes and physical objects. Historians and cultural theorists detect such waves.  In culture or economies, as in simulatrix, those waves' natural sharpening via quantization must necessarily discard the entropy and autonomy of its components, that is to say discard the self-determination and often pleasure of the cloud of human beings who inadvertently propagate those parasitic patterns.

It is easy to imagine how society might refuse, by benign neglect, to accept and keep such an inescapable indictment of its very structure. It is even harder to imagine how, if it did accept that truth, it could fail to change.

[31]There are downsides to discovering that one's unconscious existence is informationally far richer than usually thought.  One downside is realizing that the very hyper-quantized and abstract abilities that we symbol-manipulators prize in fact represent a very real distance from our most authentic selves; some of us have been ignoring 99.9% of our abilities for

most of our lives. Another downside is the pain of being unable to express that richer self in quantized ways. For example: *Imagine a little girl imprisoned in a cell, wide eyes crying through the bars. She loves to move wonderfully and gracefully, and to dance around her cell, but no one can see her there…all she can do is yell in vain around the corner. Her cage's bars are words.*

[32] A translation glossary between current western spiritual terms (which I affectionately dub "woo-woo") and classical physics might contain pairings like these:
  "Vibrations": High-bandwidth mechanical signals, or their sensory reflections
  "Doing work": psychological effort multiplied by progress in that direction
  "Authentic, vulnerable": A nervous system in a vibrationally receptive state, often accomplished by relaxing the pelvic floor or yogic breathing.
  "Faith, Universe providing": Adopting prior assumptions that our unconscious minds will operate and interact properly without interference.
  "Showing up, being present": physical co-presence in 3-D space
  "Energy, energetics": The visualizations of vibratory connections inside and between people (correlating tightly with locations of actual energy expenditure by muscles and the brain).
  "Autonomy, free will": Any entity trying to determine its effect on the world must effectively flip coins to make a secret code (a cryptographic *private key*). If it ever sees elements of that key in its sensory stream, it has information about how it affected the world.

[33] Ensnarled with thousand-dimensional recurrent feedback, the real-time system on its own must be chaotic. But chaos can be controlled, both at the millisecond level of postural stability and the years-long scale of interpersonal strategy. Real-time control requires attractors based on generic metrics of mechanical stability, while at longer timescales it requires motor attractor basins corresponding to the basic emotional states (e.g. compassion, willpower, surprise, or curiosity), each triggered by specific trans-sensory statistical metrics (respectively but approximately: vibratory synchrony, effort, temporal derivatives, or predicted information gain[33]). A well-tuned set of trans-sensory instincts ought to let a hominid run, gesture, vocalize, and eat.

[34] Digital-computer instabilities include endless loops, recursive lookups, unreachable code, non-terminating programs, memory overflows, and viruses.

[35] Resource tradeoffs include self-auditing vs. execution; short-term survival pleasures vs. long-term calibration pleasures simplified calibration stimuli (pure tones and colors) vs. complex ones like multiscale shapes and narratives; spatial vs. interpersonal calculations; episodic focus vs. "now."

[36] In the typical conditions for life (a sunlit planet in mass and radiative equilibrium), self-replication of any kind, whether via DNA or other magic molecule, actually makes the hallowed Second Law run *backwards*. Reversing entropy's direction then makes the default end-game not heat-death but the zero state: the utter elimination of diversity, autonomy, or noise. In other words, extrapolating the viral growth of any kind of copying means that overall quantifiable diversity, from genetic up to ethnic and economic diversity, will inevitably compress and standardize over time (as has been actually been observed over geologic and historic time) unless specifically counteracted. Even if the winning DNA is as fancy as ours, when rivals die, diversity goes down. (Such evolution is never neutral--*selective replication causes bias*—because the bias is in fact the principle of selection).
  The big brains of hyper-communicative, hyper-tactile homonids learned to quantize: first vibratory patterns (like gaits and songs), then representation (word meaning). Finally, perhaps because of language, our ancestors quantized physical objects. That is, is in computational terms they quantized the physical memory, or in humanistic terms they quantized the "material culture" created by "productivity." Both terms refer to information transmitted by self-replicating technology, which by its physical persistence ossified categories far more durably than any individual's fallible memory ever could.  These were the primal "persuasive technologies," such as spears and loincloths, which evolved to replicate themselves while inhibiting both their creators' immediate pains and their distractingly unproductive pleasures.  In effect productivity reduced the range of available signals along the very axes which matter most to people, and by doing so necessary sends people into deep decalibration, relative to innate sensory capacities. This principle suggests  that each successive generation has less experience of our sensory birthright. While every generation seems to have lamented the shuttered childhoods of their grandchildren, all those generations might have been correct.

   The trend of inhibiting proximity communication continues today with baggy clothing, sub-nuclear families, solitary office cubicles,  and hyper-salient but intrinsically low-bandwidth text messages and multi-player online video games.   The obvious singularity is communicating only with  emoticons (vs. its antithesis, ecstatic naked dance).

   [37] A brain learns the same way no matter what the body's size or number of limbs. A brain is born as a learning organ; the fact that it can learn (and re-learn) to control so many different kinds of joints with simple, general principles means at any stage of life and any skeletal structure, improvements can be made and physical alignment (and thus happiness) increased.  Even if the only addressable  muscles are on the face, the brain still wants to feel in control of them and to express itself. Give a brain some adhesions to release and muscles to control under natural circumstances, it will be happy.

   [38] The very imprecision of the invented term "force-geodesic" should indicate its granularity. Here is the concept: suppose the skeleton is modeled as a "tensegrity" structure of cables and struts (having struts where bones would be).  The million muscles can be viewed as a continuous 3-D mat or substance, through which co-activated micro-cables transiently align as "virtual muscles" so that their collective tension well-approximates a single cable along the line of tension.   In this scheme what matters is the exact path the central tension arc, and the synchronized vibrations it. The arc should take the form of a geodesic or most direct route, like a great circle on the earth or path of minimum curvature.  Activating muscles along these force-geodesics would be useful, if only because a force-geodesic is a continuously-moving 3-D ribbon in 3-D space, and thus has the same form as an ideal brain's representational substance.

     If a force-geodesic is proved to be some kind of mechanical optimum, it may exhibit the same kind of phase-synchrony exhibited by many wave-based systems near local maxima and minima, as happens in the calculus of variations and the principle of least action. Such coherence would be a very useful target for an ideal brain in seeking ideal postures. In particular, longitudinal vibrations along and symmetrically around the spine might enable it to naturally "feel" its ideal geodesic shape.

   [39] There is remarkable unity among the somatically aware (e.g. yoginis, martial artists, dancers) that pelvic coordination (colloquially: *inner core*; strictly: inner spinal muscles like *multifidus, transverse abdominus,* and *psoas*) is enormously important to good posture and pleasurable biomechanics.  So any native program for upright posture probably harnesses internal pleasure circuits there.  Furthermore, engorgement and its pressurization of the genitalia, regardless of cause, affects not only the vibro-mechanical structure of the pelvic region, not only the animal's appearance, not only (in males) the vibratory conduit from

sensitive external tissue to proprioception-enabling musculature, not only the relative salience of that tissue compared to other body regions, but most importantly the animal's enthusiasm and sociability. All factors must play pivotal parts in socio-sensory dynamics.

[40] What might be the topology of the perfect biped body-map? For simplicity, consider a cylindrical body, full of muscle with a spine down the middle. While the synchronization signals to those muscles might also be a cylinder, those signals are not themselves limited to the same topology; they could, for example, virtually connect one end of the spine to another to enable a looped circuit (mathematically, a hyper-torus), which might facilitate balance and long-range spinal coordination. The kundalini-style visualization for such activation might be an intensely powerful or pleasurable sensation of thread of heat and light threading through or next to the spine from one vibro-mechanically salient region of the skull (e.g. occiput, crown, brow, nose) to a corresponding one near the pelvis (tailbone, perineum, urethra).

[41] Other complementary release techniques include vibration-sources besides the body's local muscles. For example, acupuncture needles transduce body tremor, and energetic healing employs the high-frequency vibrations of a well-trained fellow human.

[42] Fascia on the skull seem to act both anatomically and perceptually as a vibratory antenna, because the closely-spaced muscles of the center spine spread into them. Having better-separated vibratory fibers packed solidly against acoustically-coupled bone makes it easier to distinguish vibrations between fibers. Thus the *skull* increases proprioceptive *spinal* resolution. In my own experience (and that of others), myofascial self-massage on the skull improves proprioception throughout the body.

[43] Knowing that quick-and-dirty communication is an optimum can be seen as giving permission to highly interactive pairs like married couples to accept one another's mistakes as part of the optimum overall strategy. Other signal-processing observations operate in the same vein. For instance, knowing that moods among nearby individuals tend to automatically synchronize, and knowing that each individual has legitimately different computational or narrative strategies, can be a neutral, blame-free way of explaining dysfunctional dyad dynamics.

[44] I personally feel as confident in predicting simulatrix's basic function as Pauli might have felt about neutrinos, or Dirac about the positron.

[45] Computer scientists recognize that while data in digital channels can be copied indefinitely, trust (or authentication) meta-data cannot be copied, but must be renewed for each transaction.

[46] The concept of "natural statistics" is an outgrowth of "natural scenes." For upright homonids, presumably similar to Bonobos except with better posture, such environments would have been highly tactile, interactive, intimate, and by current standards uncomfortable, because those homonids lived entirely outdoors unclothed and presumably communicated with expressions, body contact, and vocalizations instead of words.

[47] The "hypernormal" stimuli known to psychologists (as outlined in the popularization *Kludge*, by Gary Marcus) are for humans usually also hyper-salient, that is statistically unrepresentative of the environment for which the human nervous system evolved, but narrowed by tradition and commercial pressures precisely for their ability to grab our attention.

[48] A cursory summary of human cognitive biases (e.g. http://rationalwiki.org/wiki/List_of_cognitive_biases) includes the necessary assumptions that one's nervous system works, but additionally shows a pronounced tilt toward the same assumptions required for 3-D estimation, like continuous interpolation and extrapolation. Furthermore, a fundamental 3-D processing strategy in brains might leave traces of three-dimensionality in other aspects of cognition, such as our tendency toward groups, rules, and laws of three.

[49] Author's personal note (*written as if to my recently-adult children*) :

*Dear Sophia and Benjamin,*

*Here's the best advice I can possibly give you, and I'm using your real names to show I mean it. And using my real work, which you've seen me banging at for years.*

*Forget career advice, this is way better: the neural theory of happiness. It all boils down to two ideas. First, pleasure is the information your body and brain need most, so enjoying yourself in natural ways (like pre-verbal "cave men") and recalibrating your nervous system are practically the same thing. Second, you need to be near and visible to other people, and they to you, ideally people you have known awhile, so you can practice the physiological trust need.*

*You are both adult now, but I remember how you both loved running around outside and building forts, or climbing rocks, playing with snow. That's what I mean.*

*So what do I advise? Practice copious touch, the only output which humans still do better than machines, by far, so far. Connect with other people in person; look them in the eye, hug, dance. Those are the only signals a body really trusts (every kind of digital intermediation, including phones, has been designed to compress-out nuance and minimize your interpersonal bandwidth for profit's sake, and they're frustrating as hell. The kind of trust you need can't be wired.)*

*Don't settle for less than real-live 3-D experience (and try to visit natural places, which weren't crafted to be seen). Camp out. Take hikes.*

*Don't be afraid of discomfort. In fact, seek it out to control it. Targeted discomfort is the other necessary pole of recalibration, which we don't get directly from Nature any more. It's your body's bulwark against nonstop sweets and pillows.*

*Don't "exercise"; that's boring and repetitive. Our bodies evolved for dance and song, whatever feels good. Try continuous fun motion, not reps.*

*Take care of your spine. Get a pro to help you straighten it if necessary. If you can coax each joint to move under its own power, the whole thing will reassemble and life will be good. Really. Aches and pains are merely telling you which muscles can't yet control themselves.*

*Chase serious sensory pleasure, including all the yummy warmth mis-labeled "sex." That's a kind of calibration too.*

*Find solitude and silence—desert camping, anyone?-- to notice subtler experiences. Our ancestors camped outside for a couple million years, with nearly no noise.*

*Don't waste time on memory. "Now" is our native hyper-retinal display, and the only thing that's really real.*

*Most of all, recognize how you resonate with lovers and friends. You can't help it when your moods synch up like yawns (or arguments!). Try to cut them slack; 99.9% of what you both do is unconscious, so toes get stepped on from time to time.*

*Pass this message on to friends, in person. On paper if possible, which is easier to read and scribble on. These ideas apply to all of us, every human, because we have the same tear-ducts, throats, bellies, and butts. Our happiest and most natural existence really is the essence of simplicity: beneath the words and thoughts, we're really all the same simple social upright creatures wanting to be together and feel amazing in our skins.*

*Love, Dad*